\documentclass[aps, 
twocolumn,review,
groupedaddress ,bibnotes]{revtex4-2}

\usepackage{multirow}
\usepackage{array}
\usepackage{tikz}
\usetikzlibrary{shapes.geometric, positioning, arrows.meta, calc}
\usepackage[top= 1in, bottom= 1in, left=.78in, right=.77in]{geometry}
\usepackage{graphicx}
 \usepackage{amsthm}
 \theoremstyle{plain}

 \usepackage{graphicx}
\usepackage{mathrsfs}
\usepackage{amssymb}
\usepackage{amsmath}
\usepackage{amsfonts,bm}
\usepackage[  colorlinks=true,
    linkcolor=blue,
    citecolor=blue]{hyperref}

\usepackage{pgffor}
\foreach \x in {A,...,Z}{
  \expandafter\xdef\csname  b\x\endcsname{\noexpand\mathbb{\x}}
}
\foreach \x in {A,...,Z}{
  \expandafter\xdef\csname c\x\endcsname{\noexpand\mathcal{\x}}
}
\foreach \x in {A,...,Z}{
  \expandafter\xdef\csname s\x\endcsname{\noexpand\mathscr{\x}}
}
\foreach \x in {A,...,Z}{
  \expandafter\xdef\csname sf\x\endcsname{\noexpand\mathsf{\x}}
}
\foreach \x in {a,...,z}{
  \expandafter\xdef\csname sf\x\endcsname{\noexpand\mathsf{\x}}
}
\foreach \x in {A,...,Z}{
  \expandafter\xdef\csname  fk\x\endcsname{\noexpand\frak{\x}}
}
\foreach \x in {a,...,z}{
  \expandafter\xdef\csname  fk\x\endcsname{\noexpand\frak{\x}}
}

  \newcommand{\Tr}{\text{Tr}}

\newcommand{\be}{\begin{equation}}
\newcommand{\ee}{\end{equation}}
\newcommand{\bpm}{\begin{pmatrix}}
\newcommand{\epm}{\end{pmatrix}}

\newcommand{\PBK}[1]{\ensuremath{\begin{pmatrix}#1\end{pmatrix}}}

\newcommand{\beqn}{\begin{eqnarray}}
\newcommand{\eeqn}{\end{eqnarray}}

\newcommand{\p}{\partial}

\begin{document}

\title{Emergent $\cN=3$   and Frozen $\cN=2$ SCFTs     from   $\cN=1$ Gauge Theory
}
\author{Hongliang Jiang}
 \email[]{jianghongliang@fudan.edu.cn}
\affiliation{Center for Mathematics and Interdisciplinary Sciences, Fudan University, Shanghai 200433,
China}
\affiliation{Shanghai Institute for Mathematics and Interdisciplinary Sciences (SIMIS), Shanghai 200433,
China}
 
%
%
%
 
\begin{abstract}

Four-dimensional $\cN=3$ superconformal field theories (SCFTs) are an exotic class of quantum field theories that admit no conventional Lagrangian description. In this Letter, we construct an $\cN=1$ supersymmetric gauge theory and argue that its infrared fixed point exhibits supersymmetry enhancement to an $\cN=3$ SCFT. The ultraviolet theory has gauge group $ {SU}(2)^3$, a set of chiral multiplets, and a suitable superpotential. By performing $a$-maximization, we obtain infrared central charges $a=c=5/4$, matching those of the rank-one $\cN=3$ $\bZ_3$ S-fold SCFT. The ultraviolet gauge-theory description allows us to compute the full superconformal index with four independent fugacities.
We then examine various limits of the full index and further propose a closed-form expression for the Macdonald index, reproducing all known results for the proposed \(\cN=3\) theory.  More remarkably, we argue that further gauging the $\mathcal{N}=3$ theory yields the frozen rank-one $\mathcal{N}=2$ SCFT conjectured from Seiberg--Witten geometry, providing strong evidence for its existence and unlocking its previously inaccessible full superconformal index.
Our construction thus opens a systematic route to exploring the refined protected structure of these exotic, intrinsically strongly coupled   theories using conventional $\cN=1$ Lagrangian techniques.

\end{abstract}

\maketitle

\section{Introduction} 
Understanding the non-perturbative dynamics of quantum field theory (QFT) is a central challenge in fundamental theoretical physics. These dynamics not only govern the infrared behavior of QFTs but also give rise to exotic phenomena and emergent structures. Strong-coupling effects, however,  often render conventional perturbative methods inadequate; more fundamentally, some theories admit no conventional Lagrangian description at all. Developing new tools to access these strongly coupled regimes is therefore essential to uncovering the rich landscape of non-perturbative QFT.

Four-dimensional $\cN=3$ superconformal field theories (SCFTs) form an exotic class of intrinsically strongly coupled QFTs and provide an especially intriguing setting for these questions. With twelve Poincar\'e supercharges, they lie between the extensively studied $\cN=2$ theories and $\cN=4$ super-Yang--Mills theory, yet remain comparatively elusive. The obstacle to a conventional Lagrangian description is structural: a weakly coupled theory with $\cN=3$ supersymmetry necessarily enjoys $\cN=4$ supersymmetry, so genuine $\cN=3$ SCFTs must be intrinsically strongly coupled~\cite{Aharony:2015oyb}. This automatic supersymmetry enhancement long fueled the belief that genuine $\cN=3$ theories might not exist at all. That expectation, however, was overturned in 2015 by the string-theoretic S-fold construction \cite{Garcia-Etxebarria:2015wns}, thus bypassing  the weak-coupling obstruction of direct Lagrangians. 
%
Their non-Lagrangian nature echoes that of $\cN=2$ Argyres--Douglas theories, discovered two decades earlier~\cite{Argyres:1995jj}, whose fractional Coulomb-branch scaling dimensions likewise obstruct a conventional $\cN=2$ Lagrangian description.

Since their discovery, $\cN=3$ SCFTs have been investigated through string theory, Seiberg--Witten geometry, vertex operator algebra, conformal bootstrap and holography, revealing rich structures in their moduli spaces and   spectra. Nevertheless, a basic question remains: can these exotic theories be realized using the well-developed tools of conventional gauge theory, and can such an approach provide systematic access to their refined observables?

\begin{figure}[htbp]
    \centering
     \includegraphics[width=0.5\textwidth]{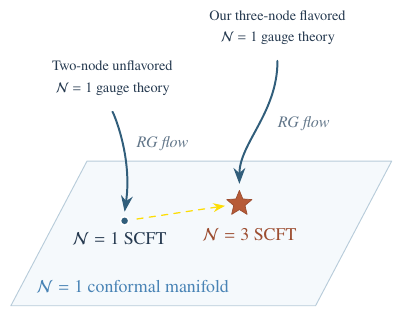}
    \caption{IR fixed points of       our  three-node gauge  theory and  the two-node    theory in   \cite{Zafrir:2020epd}, which are connected by an $\cN=1$ preserving marginal deformation.}
    \label{figrgflow}
\end{figure}

In this Letter, we provide an affirmative answer for the simplest $\cN=3$ SCFT. We construct a simple $\cN=1$ quiver gauge theory and argue that its infrared fixed point exhibits enhanced $\cN=3$ supersymmetry, realizing the rank-one $\bZ_3$ S-fold SCFT. Our construction thus provides an indirect Lagrangian description of this intrinsically strongly coupled theory. The strategy of accessing non-Lagrangian fixed points through supersymmetry enhancement in $\cN=1$ gauge theories was pioneered for   $\cN=2$ Minahan-Nemeschansky  $E_6$ theory in \cite{Gadde:2015xta} and   Argyres--Douglas theories in  \cite{Maruyoshi:2016tqk}. Although this strategy has been extensively developed in the $\cN=2$ setting, 
 realizing a flow directly to an enhanced $\cN=3$ fixed point presents a distinct challenge.  An important earlier proposal was made in  \cite{Zafrir:2020epd}, where the interacting sector of an $\cN=1$ gauge theory was argued to flow generically to an $\cN=1$ point on the conformal manifold containing the $\cN=3$ SCFT, rather than directly to the enhanced $\cN=3$ point itself. See Fig.~\ref{figrgflow} for illustration.

A distinguishing feature of our construction is its manifest rank-two flavor symmetry, in addition to the $\cN=1$ $U(1)_R$ symmetry. Under the proposed flow, the two flavor Cartan generators, together with the ultraviolet $U(1)_R$ generator, span the full Cartan subalgebra of the enhanced infrared $\cN=3$ $R$-symmetry.

Our proposal is supported by the agreement between the infrared central charges obtained through $a$-maximization and those of the proposed $\cN=3$ SCFT. More importantly, the ultraviolet gauge-theory description allows us to express the full superconformal index, with four independent fugacities, as a threefold contour integral. This provides the first fully refined expression for the superconformal index of a genuine  and intrinsic $\cN=3$ SCFT \footnote{\cite{Bourton:2018jwb} also computed the full index of an $\cN=3$ SCFT, which, however, arises from discrete gauging of an $\cN=4$ theory and is therefore not intrinsically $\cN=3$.}. At the first few nontrivial orders, the index expansion organizes into contributions from $\cN=3$ supermultiplets, providing a direct test of the proposed supersymmetry enhancement. By evaluating various limits through perturbative and analytic calculations, we obtain the Coulomb, Hall--Littlewood, Schur, and Macdonald indices, reproducing all known results in the literature for these limits.

Remarkably, an additional gauging of the $\cN=3$ SCFT leads to an infrared fixed point that we propose to identify with the frozen rank-one $\cN=2$ SCFT $ { {IV}^{*}_{Q=1}}$,  which was previously    conjectured from Seiberg--Witten geometry \cite{Argyres:2016xmc}. Our construction not only confirms the existence of such a theory, but also yields an explicit prescription for computing its previously inaccessible full superconformal index.
  

  \section{The  $\cN=1$    gauge theory}

\begin{figure}[t] 
    \centering
    \resizebox{\columnwidth}{!}{%
    \begin{tikzpicture}[
        x=1cm, y=0.9cm,
        gauge/.style={circle, draw, thick, minimum size=1.1cm, inner sep=1pt, font=\small},
        box/.style={rectangle, draw, thick, minimum width=0.7cm, minimum height=0.7cm, inner sep=2pt, font=\small},
        line/.style={thick, -} 
    ]

        \node[gauge] (SU2C) at (0,0) {\(SU(2)_C\)};
        \node[gauge] (SU2L) at (-1.9,-3.5) {\(SU(2)_L\)}; 
        \node[gauge] (SU2R) at (1.9,-3.5) {\(SU(2)_R\)};  
        
        \node[box] (BoxTop) at (0, 2.2) {2};       
        \node[box] (BoxL) at (-1.9, -5.7) {2};    
        \node[box] (BoxR) at (1.9, -5.7) {2};     
        
        \coordinate (V) at (0, -1.75);
        
        \draw[line] (BoxTop) -- (SU2C) node[midway, right, font=\small] {\(C_\pm\)};
        
        \draw[line] (SU2C) -- (SU2L) 
             node[pos=0.6,  above left, font=\small] {\( \color{blue} M\)}
            node[pos=0.6,rotate=60, font=\Large] {\(\color{blue} {\times}\)}
            node[pos=0.3, above left, font=\small] {\(X\)}; 
            
        \draw[line] (SU2C) -- (SU2R)  node[pos=0.6,  above right, font=\small] {\(\color{blue}N\)}
            node[pos=0.6,rotate=120, font=\Large] {\(\color{blue} {\times}\)}
            node[pos=0.3, above right, font=\small] {\( Y\)}; 
        
        \draw[line] (SU2C) -- (V) node[midway, below,yshift=-17pt, font=\small] {\(Q\)};
        \draw[line] (V) -- (SU2L);
        \draw[line] (V) -- (SU2R);
        
        \draw[line] (BoxL) -- (SU2L) node[midway, right, font=\small, xshift=2pt] {\(L_\pm\)};
        \draw[line] (BoxR) -- (SU2R) node[midway, left, font=\small, xshift=-2pt] {\(R_\pm\)};
        
        \draw[thick, -] (SU2L) edge [loop left, distance=1.18cm, looseness=1.3, -] 
            node[below, yshift=-0.35cm, xshift=0.5cm, font=\small] {\(J\)} (SU2L);
        
    \end{tikzpicture}%
    }
    \caption{Quiver diagram for the $\cN=1$ \(SU(2)^3\) gauge theory and chiral multiplet content. Circle nodes are gauge nodes, and boxes indicate the multiplicity of fundamental chiral multiplets. 
    Each link represents a (bi-) fundamental chiral multiplet, the loop $J$ is the adjoint chiral, and the trivalent junction $Q$ represents a tri-fundamental. The cross denotes a flipping field for the mesonic operator.   
    }
    \label{figquiver}
\end{figure}
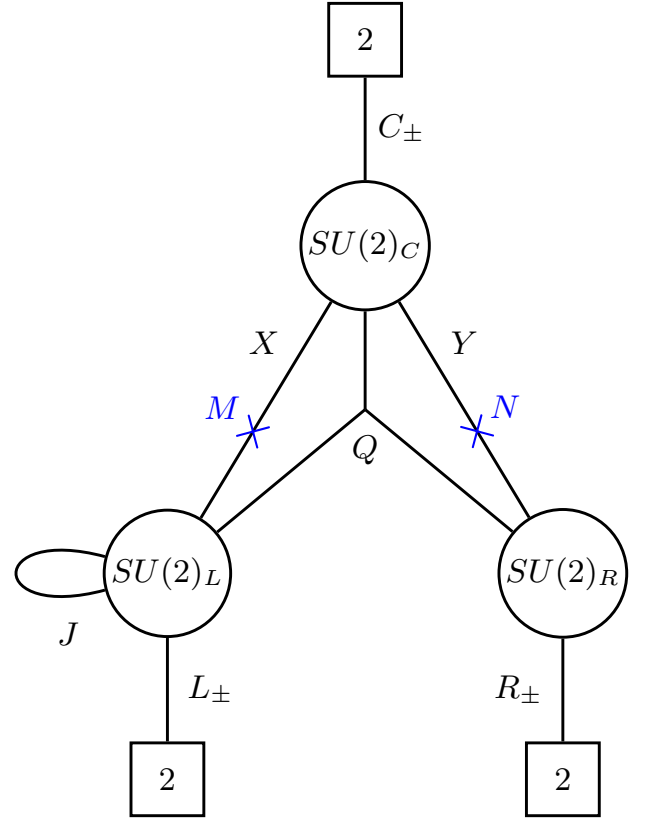


\begin{table}[h]
\centering
\renewcommand{\arraystretch}{1.85}  
\begin{tabular}{c | c | >{\hspace{.81em}}  r <{\hspace{.9em}}|@{\hspace{.9em}} r@{\hspace{1.875em}} r<{\hspace{1em}}} 
\hline
Matter &  \scalebox{0.9}{\({SU}(2)_L \!\times\! {SU}(2)_C \!\times\! {SU}(2)_R\)} & \(R\) & \(V\) & \(H\) \\
\hline
\(J\) & \((\mathbf{3},\mathbf{1},\mathbf{1})\) & \(\frac{2}{3}\) & \(-2\) & \(0\) \\
\hline
\(Q\) & \((\mathbf{2},\mathbf{2},\mathbf{2})\) & \(\frac{2}{3}\) & \(1\) & \(0\) \\
\hline
\(X\) & \((\mathbf{2},\mathbf{2},\mathbf{1})\) & \(\frac{1}{3}\) & \(\frac{1}{2}\) & \(-\frac{1}{2}\) \\
\hline
\(Y\) & \((\mathbf{1},\mathbf{2},\mathbf{2})\) & \(0\) & \(-\frac{3}{2}\) & \(\frac{1}{2}\) \\
\hline
\(L_+\) & \multirow{2}{*}{\((\mathbf{2},\mathbf{1},\mathbf{1})\)} & \(\frac{4}{3}\) & \(\frac{1}{2}\) & \(-\frac{1}{2}\) \\
\cline{1-1} \cline{3-5}
\(L_-\) & & \(\frac{2}{3}\) & \(\frac{5}{2}\) & \(\frac{3}{2}\) \\
\hline
\(C_+\) & \multirow{2}{*}{\((\mathbf{1},\mathbf{2},\mathbf{1})\)} & \(1\) & \(-3\) & \(-1\) \\
\cline{1-1} \cline{3-5}
\(C_-\) & & \(\frac{5}{3}\) & \(1\) & \(1\) \\
\hline
\(R_+\) & \multirow{2}{*}{\((\mathbf{1},\mathbf{1},\mathbf{2})\)} & \(1\) & \(-\frac{3}{2}\) & \(\frac{1}{2}\) \\
\cline{1-1} \cline{3-5}
\(R_-\) & & \(\frac{1}{3}\) & \(\frac{1}{2}\) & \(-\frac{3}{2}\) \\
\hline
\(M\) & \((\mathbf{1},\mathbf{1},\mathbf{1})\) & \(\frac{4}{3}\) & \(-1\) & \(1\) \\
\hline
\(N\) & \((\mathbf{1},\mathbf{1},\mathbf{1})\) & \(2\) & \(3\) & \(-1\) \\
\hline
\end{tabular} 
\caption{Chiral multiplets, their  gauge representations, and charges under  $R$ and flavor symmetries.}
\label{chimlts}
\end{table}

 The $\cN=1$ supersymmetric gauge theory has gauge group $SU(2) ^3$ and is described by the quiver  in Figure~\ref{figquiver}. There are 12  species of chiral multiplets $\Phi_i$ whose gauge representations  are given in Table~\ref{chimlts}.   
The theory also has a  superpotential  
  \beqn\label{Wsup}
\cW&=&  h_1 XQR_+ + h_2 YQL_+ + h_3 L_- X C_+ + h_4 R_- Y C_- 
\nonumber\\
  & &
 +h_5 \, JQQ  + h_6 \, M \det X + h_7\, N \det Y , \qquad
  \eeqn
  where all the coefficients $h_i$ are non-zero, and each operator is the unique gauge singlet constructed out of the chiral multiplets.  For example,  $M \det X\equiv\frac12 M X^{am}X^{bn} \epsilon_{ab}\epsilon_{mn}$, and $JQQ\equiv\epsilon_{ab} \epsilon_{rs} J_{(mn)} Q^{ arm}Q^{ bsn}$, where $ a,m,r\cdots\in\{1,2\}$ are the fundamental indices of various $SU(2)$ gauge factors.
  $M,N$ are  gauge singlets, which are  flipping fields   whose $F$-term equations remove the    operators $\det X,\det Y$.

The cancellation of the mixed $R$–gauge–gauge ABJ anomaly  for each gauge factor imposes 
\be\label{ABJR}
T(\text{adj})+\sum_{i=1}^{12}\cN_A(\Phi_i) T ( \bm \Phi _i ^A )(R _{\Phi_i}-1)=0,\quad A=L,C,R,
\ee
where $ \bm \Phi _i ^A $ is the  representation that $\Phi_i$ transforms under gauge group $SU(2)_A$, and $T$ is the Dynkin index with  
$T(\bm 1) =0, T(\bm 2)=\frac12, T(\text{adj})=T(\bm 3)=2$. In addition, $\cN_A(\Phi_i)=\dim \Phi_i/\dim_A \Phi_i$ is the effective multiplicity seen from $SU(2)_A$ and $\dim\Phi_i= \dim_L \Phi_i \cdot\dim_C  \Phi_i \cdot\dim_R  \Phi_i$, where $\dim_A \Phi_i$ is the dimension of representation $ \bm \Phi _i ^A$. 

Similarly, the non-$R$ flavor symmetries, which commute with supercharges, should satisfy
\be\label{ABJF}
 \sum_{\Phi_i} \cN_A(\Phi_i) T ( \bm \Phi _i ^A )F _{\Phi_i} =0,\qquad A=L,C,R.
\ee
Note that   $R _{\Phi_i}+\alpha F _{\Phi_i}$ is still a valid anomaly free candidate $R$ symmetry; in particular it satisfies \eqref{ABJR}.

 The superpotential has $U(1)_R$ charge 2 and is neutral under flavor symmetries.
Given the superpotential
$\cW=\sum_{ijk}c_{ijk}\Phi_i\Phi_j\Phi_k$, the condition $ R(\cW)=2$ imposes $R_{\Phi_i}+R_{\Phi_j}+R_{\Phi_k}=2$ for non-zero $c_{ijk}$. 
The 12 chiral multiplets yield $U(1)^{12}$ symmetries, and the 7 terms in the superpotential \eqref{Wsup} give  7 constraints, yielding 5 independent $U(1)$.  The condition \eqref{ABJR} gives three further constraints, which implies that the anomaly free $R$ symmetry has 2 free parameters,    corresponding to the rank of flavor symmetry that can mix with $R$ symmetry.   Up to  a change of basis, these conditions determine  the charges in the last three columns of Table~\ref{chimlts}.

This enables us to write the trial $ U(1)_R$ symmetry as 
\be\label{RVH}
\cR=R+\alpha V+\beta H,
\ee
where $R$ is any reference anomaly-free $ R$-symmetry and $V,H$ are the  two extra flavor symmetries. The exact $ U(1)_R$ symmetry  at the infrared fixed point   can be determined by $a$-maximization \cite{Intriligator:2003jj}. 
The $a,c$ central charges are given by
\be\label{acR}
a = \frac{3}{32} \left( 3 \operatorname{Tr} \cR^3 - \operatorname{Tr} \cR \right), 
\quad 
c  = \frac{1}{32} \left( 9 \operatorname{Tr} \cR^3 - 5 \operatorname{Tr} \cR \right).
\ee
For this purpose, we need to compute $\Tr \cR, \Tr\cR^3$, 
where the trace is taken over all  the chiral fermions from both the chiral and vector multiplets. 
Note that each chiral multiplet of charge $R_\Phi$  contains a fermion  of charge $R_\Phi-1$, while the gaugino in each  vector multiplet has charge $R_A=1$ and $V_A=H_A=0$ for each $SU(2)_ {A=L,C,R}$.
Then using  \eqref{RVH} we find
\be
\Tr \cR=\sum_\Phi \dim \Phi  \cdot (\cR_\Phi-1)+\sum_A \dim \fks\fku(2)_A\cdot 1=0,
\ee
and
 \beqn
\Tr \cR^3&=&\sum_\Phi \dim \Phi  \cdot (\cR_\Phi-1)^3+\sum_A \dim \fks\fku(2)_A\cdot 1^3
\nonumber\\& =&
  \frac{10}{9}\,( 1-3\alpha )\left(4+12\alpha+9\alpha^2-9\beta^2\right).\qquad
\eeqn
 
With these results, we can compute the trial central charges using \eqref{acR}   
\be
a _\text{trial} = c_\text{trial} = \frac{5}{4} - \frac{135}{16}\alpha^2 - \frac{45}{16}\beta^2 - \frac{135}{16}\alpha^3 + \frac{135}{16}\alpha\beta^2.
\ee

The superconformal $R$ symmetry is determined by the local maximum of the trial $a$ central charge, which must satisfy 
\be
\p_\alpha a_\text{trial}=\p_\beta a_\text{trial}= 0, \qquad  \PBK{ \p_\alpha ^2 a_\text{trial} & \p_\alpha \p_\beta a_\text{trial} \\ \p_\beta\p_\alpha a_\text{trial} & \p_\beta ^2 a_\text{trial}} \prec 0.
\ee
 These conditions uniquely select  $\alpha=\beta=0$.  So $R$ is  exactly  the superconformal $U(1)_R$ symmetry at the IR fixed point. 
As a result, the IR superconformal point has central charges 
\be
a=c=\frac{5}{4}.
\ee
These central charges coincide with those of  $\cN=3$    $\bZ_3$ S-fold SCFT, which is arguably the simplest   $\cN=3$ SCFT. 
The agreement of central charges motivates us to propose that   our $\cN=1$ supersymmetric gauge theory flows to the IR fixed point with enhanced $\cN=3$ supersymmetry, yielding the $\cN=3$  $\bZ_3$ S-fold SCFT. We will provide further non-trivial evidence  to   justify the proposal below.

  \section{Superconformal index}
We now study the full index from the gauge theory, as well as its various limits.

\subsection{Full index from the gauge integral}
We have two   $U(1)$ flavor symmetries, so we introduce two flavor fugacities $\sigma, \xi$ and study the flavored  index as follows:
\be\label{N1index}
\mathcal{I}  =
\operatorname{Tr}(-1)^F
p^{j_1+j_2+\frac R2}
q^{-j_1+j_2+\frac R2}
\sigma^{V}\xi^{H},
\ee
where the trace is taken over operators satisfying the condition $E=2j_2+\frac32R$.    

The standard prescription  for computing  the index gives the following integral \cite{Gadde:2020yah}

 \begin{align}
&\mathcal{I}   = \left(\frac{\kappa}{2}\right)^3 \oint \frac{dx}{2\pi i x} \frac{dy}{2\pi i y} \frac{dz}{2\pi i z} \;
\frac{1}{ 
 \Gamma_e(x^{\pm 2}) \Gamma_e(y^{\pm 2})  \Gamma_e(z^{\pm 2  } )}
\nonumber \\ 
 &  \times  
  \Gamma_e\!\left((pq)^{\frac{1}{3}}\sigma^{-2}\right)
\Gamma_e\!\left((pq)^{\frac{1}{3}}\sigma^{-2} z^{\pm 2}\right) 
\nonumber\\ 
  &  \times 
\Gamma_e\!\left((pq)^{\frac{1}{3}}\sigma\, z^{\pm 1}y^{\pm 1}x^{\pm 1}\right)
 \Gamma_e\!\left((pq)^{\frac{1}{6}}\sigma^{1/2}\xi^{-1/2} z^{\pm 1}y^{\pm 1}\right)
\nonumber  \\ &  \times
\Gamma_e\!\left(\sigma^{-3/2}\xi^{1/2} y^{\pm 1}x^{\pm 1}\right) 
 \Gamma_e\!\left((pq)^{\frac{2}{3}}\sigma^{1/2}\xi^{-1/2} z^{\pm 1}\right)
\nonumber \\ &  \times 
\Gamma_e\!\left((pq)^{\frac{1}{3}}\sigma^{5/2}\xi^{3/2} z^{\pm 1}\right)   \Gamma_e\!\left((pq)^{\frac{1}{2}}\sigma^{-3}\xi^{-1} y^{\pm 1}\right)
\nonumber\\
&  \times
\Gamma_e\!\left((pq)^{\frac{5}{6}}\sigma\xi\, y^{\pm 1}\right) \Gamma_e\!\left((pq)^{\frac{1}{2}}\sigma^{-3/2}\xi^{1/2} x^{\pm 1}\right)
\nonumber\\
&  \times 
\Gamma_e\!\left((pq)^{\frac{1}{6}}\sigma^{1/2}\xi^{-3/2} x^{\pm 1}\right) 
\nonumber\\
&  \times 
 \Gamma_e\!\left((pq)^{\frac{2}{3}}\sigma^{-1}\xi\right)
\Gamma_e\!\left(pq\,\sigma^3\xi^{-1}\right), 
\label{Idxgauge}
\end{align}
where the weight $(pq)^{R/2} \sigma^V\xi^H$ can be read off from   Table~\ref{chimlts} and $x,y,z$ are the gauge fugacities.  We use the notation 
\be
\Gamma_e(z ) \equiv\Gamma_e(z;p,q)  = \prod_{m,n \ge 0} \frac{1 - p^{m+1} q^{n+1}/z}{1 - z p^m q^n},
\ee
and
$ (z;p)_\infty  = \prod_{m \ge 0} (1 - z p^m),  \kappa = (p;p)_\infty (q;q)_\infty$. 

It is convenient to introduce a new   set  of charges $\sfR, \sfr, \sfF$, which are related to  $R,V,H$  via
$ R   = \frac{ 1}{3}(4\sfR-2\sfr) ,  V= -\sfR - \sfr - \frac{3}{2}\sfF , H = -\sfR - \sfr + \frac{1}{2} \sfF$.      
%
  %
  Meanwhile, for later purposes, we also introduce a new set of fugacities via
  \be\label{pqst}
  p=\tau^3 \rho, \quad q=\tau^3/\rho, \quad s= \sqrt{ { \xi}/{\sigma^3}}, \quad t= {\tau^4}/{(\sigma\xi)}.\qquad
  \ee

As a result, the     superconformal index in  \eqref{N1index} can be rewritten as  
\be\label{n2indx}
\mathcal{I}(p,q,t,s) = \text{Tr}  (-1)^F p^{j_1+j_2-\sfr} q^{j_2-j_1-\sfr} t^{\sfR+\sfr} s^{\sfF},  
\ee
where the trace receives contributions from   operators satisfying $  E - 2j_2 - 2\sfR + \sfr=0$. 
 The formula \eqref{n2indx} has precisely the structure of 
 an index for $\cN=2$ SCFTs with an extra flavor symmetry $U(1)_ \sfF$. 

%
 \subsection{Perturbative expansion of the index}
 
 We now   compute the index by evaluating the gauge integrals  \eqref{Idxgauge} explicitly. 
  It is convenient to    use the fugacities $\tau,\rho,\sigma,\xi$ defined in \eqref{pqst} , and compute the index perturbatively in $\tau$ \footnote{Because of the subtlety associated with $R(Y)=0$, the contour integral should be evaluated in the chamber
$|s|=|\sqrt{\xi/\sigma^3}|<1$, with the limit $s\to1$ taken afterward if needed.
For convenience, one may rescale the fugacities as
$\xi=\tau^{1/4}\hat\xi$ and $\sigma=\tau^{-1/4}\hat\sigma$.
We then use $\tau,\rho,\hat\xi,\hat\sigma$ as independent variables and expand the integrand in $\tau$.
The integral can be evaluated order by order by taking residues at the origins of the gauge-fugacity planes, and the final result can be reexpressed in terms of $\tau,\rho,\sigma,\xi$. }.   The first few orders are given by~\footnote{We have evaluated the full index up to the order $\tau^{60}$ and we show the result up to order $\tau^{20}$ in appendix. }
   \beqn\label{idpert}
\cI&=  &1+ \frac{\chi_{\bm 2}(\xi)}{\sigma} \tau^4 - \frac{\chi_{\bm 2}(\rho)}{\sigma^2} \tau^5
  \nonumber \\  &&
   + 
  \Big[ \sigma^3 \chi_{\bm 4}(\xi) - \chi_{\bm 3}(\xi) - 1 + \sigma^{-6} \Big] \tau^6 
  \nonumber \\  &&
  + \chi_{\bm 2}(\rho) \left[ \frac{2\chi_{\bm 2}(\xi)}{\sigma} - \sigma^2 \chi_{\bm 3}(\xi) \right] \tau^7
  \nonumber \\
 &&+ \Big[ -\sigma^4 \chi_{\bm 3}(\xi) + (2\sigma - \sigma^{-5})\chi_{\bm 2}(\xi) 
+ \frac{  \chi_{\bm 3}(\xi) - \chi_{\bm 3}(\rho) - 1  }{\sigma^{ 2}} \Big] \tau^8
\nonumber \\
&&+ O(\tau^9),
\eeqn
where $\chi_{  \bm d}(w)=w^{d-1}+w^{d-3}+\cdots + w^{1-d}$ is the character of the  $  d$-dimensional  representation of $\fks\fku(2)$.
The character structure above strongly suggests an enhancement of the infrared flavor symmetry algebra to $\mathfrak{su}(2)\times\mathfrak{u}(1)$. From the $\cN=1$ perspective, this is precisely the flavor symmetry expected of an $\cN=3$ SCFT: it arises from the subalgebra of the $\cN=3$ $R$-symmetry $\mathfrak{su}(3)\times\mathfrak{u}(1)$ that commutes with the chosen $\cN=1$ superconformal subalgebra. Indeed, a genuine $\cN=3$ SCFT admits no additional continuous flavor symmetry, so these apparent $\cN=1$ flavor symmetries are naturally identified with part of the enhanced infrared $R$-symmetry.
Furthermore, one can show that the reduced index
$(1-p)(1-q)(\mathcal I - 1) $ contains $  \Big(  \sigma^{-1}\xi + \sigma^{-1}\xi^{-1} 
- \sigma^2\xi^2 - \sigma^2 - \sigma^2\xi^{-2}\Big) \tau^7 \chi_2(\rho)$, where the first two terms with positive coefficient also indicate  the supersymmetry enhancement from $\cN=1$ to $\cN=3$, following the argument in \cite{Evtikhiev:2017heo},  as explained in  \footnote{The general statement in  \cite{Evtikhiev:2017heo} is that: for an interacting theory whose reduced index is $ (1-p)(1-q)(\mathcal I-1)$, 
a positive integer coefficient \(k\) of \(\tau^7\chi_2(\rho)\) implies enhancement of supersymmetry to at least \(\mathcal N=k+1\). 
Here we use the flavor   resolved version of this statement. Also it is easy to argue that our IR theory can not have $\cN=4$ supersymmetry. 
}.

 The $\cN=3$ structure of the index becomes more transparent upon decomposing it into contributions from $\cN=3$ superconformal multiplets \cite{Cordova:2016emh}. In particular, the first few orders of \eqref{idpert} can be written as
\beqn
\cI &=& 1+\cI_{B_1\bar B_1[0;0]_{2}^{(1,1;0)}}
\nonumber\\
&&+\cI_{B_1\bar B_1[0;0]_{3}^{(3,0;6)}}
+\cI_{B_1\bar B_1[0;0]_{3}^{(0,3;-6)}}
+O(\tau^8),\qquad
\eeqn
where $1$ denotes the identity contribution, the second term corresponds to the stress-tensor multiplet, which in particular  contains the additional supercurrents required for supersymmetry enhancement to $\cN=3$  at order $\tau^7$, and the remaining conjugate pair     contains  dimension-three Coulomb and Higgs branch operators. 

Another hint of $\cN=3$ supersymmetry comes from the anomaly polynomial. 
 After properly embedding our \(R,V,H\) into the \(\mathcal{N}=3\) \(R\)-symmetry \(SU(3)_ {\cN=3}\times U(1)_ {\cN=3}\),  from  Table~\ref{chimlts} we also find  that  $ \Tr(SU(3)_ {\cN=3}^3)=\Tr(U(1)_  {\cN=3}\  SU(3)_ {\cN=3}^2)=5$, $ \Tr(U(1)_  {\cN=3}^3)=-120$,  and $ \Tr(SU(3)_ {\cN=3})= \Tr( U(1)_ {\cN=3})=0$,
which is consistent with the anomaly polynomial of \(\mathcal{N}=3\) SCFTs~\cite{Zafrir:2020epd}.

The   complementary arguments  above from various perspectives thus provide further strong evidence that our $\cN=1$ gauge theory flows to the proposed $\cN=3$ SCFT. This contrasts with the proposal of Ref.~\cite{Zafrir:2020epd}, whose ultraviolet gauge theory has no manifest continuous non-$R$ flavor symmetry and was argued to flow to a generic $\cN=1$ point on the conformal manifold containing the $\cN=3$ SCFT. Our construction instead preserves a manifest rank-two flavor symmetry, enabling a fully refined test of the proposed flow directly to the enhanced $\cN=3$ point.

  \subsection{Limits of the full index}
  
 Although the integral \eqref{Idxgauge} gives the full index, its exact evaluation is challenging. We can thus take various limits, and try to compute them analytically. In order to compare with known results of the proposed IR SCFT in the literature, we will use $p,q,s,t$ as fugacities and regard the index as an $\cN=2$ index   with   flavor symmetry \eqref{n2indx}. 
  
    \subsubsection{Coulomb   index} 
    We can introduce $u=pq/t$ and send $p,q,t\to 0$ with fixed $u,s$, which gives the Coulomb index. 
    In this limit, we find that the integral \eqref{Idxgauge}  can be evaluated analytically (see Appendix~\ref{appCBIndx}):
    \be
    \cI_C=\frac{1}{1-u^3}.
    \ee 
 This implies that, viewed as an $\cN=2$ theory, the infrared SCFT has rank one, with a Coulomb-branch operator of scaling dimension 3. Consistently, the proposed SCFT has full moduli space $\bC^3/\bZ_3$, with the identification
$
(z_0,z_+,z_-)\sim(\omega z_0,\omega z_+,\omega^{-1}z_-)$
for $  \omega=e^{2\pi i/3}$. 
The Coulomb branch is the locus $z_+=z_-=0$ and is parametrized by the invariant $ U=z_0^3$, with $\Delta_{U}=3$. The Higgs branch corresponds to $z_0=  0$, and is given by $\bC^2/\bZ_3$, with
$
(z_+,z_-)\sim(\omega z_+,\omega^{-1}z_-).
$
Its chiral ring is therefore generated by the invariants
$
W_\pm=z_\pm^3,  W_0=z_+z_-,
$
subject to the relation $W_+W_-=W_0^3$. The  corresponding Hilbert series of the ring gives the Higgs index 
   \be\label{Hgsindx}
    \cI_{\mathrm{Higgs}}(t,s)  = \frac{1-t^3}{(1-t)\left(1-t^{3/2}s^3\right)\left(1-t^{3/2}s^{-3}\right)}.
  \ee
  Note that  Higgs index can not be obtained as a special limit of full index, but the generators can be identified as 
$     W_-\sim C_+\,Y\,R_+ ,W_+ \sim \det(Q\,R_-), W_0\sim R_+R_-$ in terms of fields in Table~\ref{chimlts}. 
  
          \subsubsection{Schur   index}  
    We can also take the limit $t=q$, which gives the Schur index. Note that in this limit, the index is independent of $p$. The Schur index is very useful, because it is identified with  the normalized vacuum supercharacter of
the associated vertex operator algebra (VOA) according to the SCFT/VOA correspondence \cite{Beem:2013sza}. The VOA  of $\bZ_3$ S-fold theory has been proposed in \cite{Nishinaka:2016hbw}, which also admits a free field realization \cite{Bonetti:2018fqz}.

By manipulating   the gauge integral, we obtain the following closed-form expression \cite{Jiang2026}
\be\label{ISc}
\mathcal{I}_S
=
\frac{(s^{\pm1}q^{1/2};q)_\infty}{2(q;q)_\infty^2}
\sum_{\substack{k\ge 0\\ \epsilon=\pm 1}}
\frac{q^{3k/2}s^{-3\epsilon k}\bigl(1+s^\epsilon q^{k+1/2}\bigr)}
{\bigl(1-s^\epsilon q^{k+1/2}\bigr)\bigl(1-s^{3\epsilon}q^{3k+3/2}\bigr)},
\ee
where \((a;q)_n = \prod_{i=0}^{n-1}(1-aq^i), (a,b;q)_n = (a;q)_n(b;q)_n,  ( s^{\pm 1} x; q)_\infty\equiv (s x ,s^{-1}x; q)_\infty$.
This agrees with the flavored Schur   index computed in \cite{Guo:2026lsl} via the  flavored modular linear differential equation. 
 
       \subsubsection{Macdonald   index}  
    In the limit $p\to 0$, we get the Macdonald index, for which we propose the following closed-form   expression
    \beqn\label{Imd}
\cI_M&=& 
\frac{(q^{3/2}T \mathfrak{s}^{\pm 1}; q)_\infty}{(qT, q^2T; q)_\infty} 
\sum_{m,n\geq 0} 
q^{3mn+\frac{3}{2}(m+n)} T^{3m} \mathfrak{s}^{3(m-n)}
\nonumber\\&\times& 
  \frac{1 - T q^{m+n+1}}{1 - Tq} 
\frac{(qT, \sqrt{q} \mathfrak{s}^{-1}; q)_m}{(q, q^{3/2}T \mathfrak{s}^{-1}; q)_m} 
\frac{(qT, \sqrt{q} \mathfrak{s}; q)_n}{(q, q^{3/2}T \mathfrak{s}; q)_n},\qquad\qquad
\eeqn 
where $T=t/q,\mathfrak s = s/ {\sqrt{T}} $. 
We have verified that the formula \eqref{Imd} agrees with our gauge integral computation \eqref{Idxgauge} up to $q^{50}$, as well as the  result   in  \cite{Agarwal:2021oyl} which was computed based on the  free field realization  of VOA with   certain  filtration. Furthermore, one can show that \eqref{Imd} reduces to \eqref{ISc} in the limit $T=t/q=1$.

      \subsubsection{Hall–Littlewood   index} 
      Taking both $p,q\to 0$ with fixed $t,s$ gives the Hall-Littlewood (HL) index, which can also be obtained as the $q\to 0$ limit of the Macdonald index in   \eqref{Imd}:
\be
\mathcal{I}_{\mathrm{HL}}(t,s) 
=
\frac{1+t+t^2-(1+t)t^{3/2}s^{-1}-t^2s^2}
{(1-t^{3/2}s^3)(1-t^{3/2}s^{-3})}.
\ee
   Note that HL index is  different from Higgs index \eqref{Hgsindx} in general, due to  additional semishort-multiplet contributions beyond the Higgs chiral operators. Nevertheless, we find that  $\cI_{\mathrm{Higgs}}(t,s) = \frac{1}{3}\sum_{\ell=0}^{2}
\mathcal{I}_{\mathrm{HL}}(t,e^{2\pi i \ell/3}s)$.

%
   
   \section{Further gauging and a frozen rank-one $\cN=2$ SCFT}

The rank-one $\cN=3$ $\bZ_3$ S-fold SCFT  has flavor symmetry   $ {\fks\fku}(2)\times {\fku}(1)$ from an $\cN=1$ perspective. Adding a chiral doublet and gauging the diagonal $ {SU}(2)$ with an $\cN=1$ vector multiplet, we obtain infrared central charges $a=55/48$ and $c=25/24$ through $a$-maximization. Our full $\cN=3$ index \eqref{Idxgauge} also determines the full index of the gauged theory, which suggests a rank-one Coulomb branch generated by a dimension-three operator and a trivial Higgs branch. These characteristic properties strongly suggest the infrared fixed point is given by the  frozen rank-one $\cN=2$ SCFT $ { {IV}^{*}_{Q=1}} $, originally conjectured from Seiberg--Witten geometry \cite{Argyres:2016xmc}:
\begin{equation}
\left( \cT_{\bZ_3 \text{ S-fold}}^{\cN=3  }\oplus\bm{2}\right)
 \big/_{\cN=1} {SU}(2)_{\mathrm{diag}}
\;\xrightarrow{\mathrm{RG}}\;
\cT_{ {IV}^{*}_{Q=1}}^{\cN=2  }.
\end{equation}
A calculation of the Schur index further suggests   that the associated VOA is   the doublet algebra $\cA(4)$~\cite{Feigin:2007sp,feigin2008characters}. 
A detailed analysis will be presented in \cite{Jiang2026}.

%

\section{Conclusion}
In this Letter, we have constructed an $\cN=1$ supersymmetric gauge theory with manifest rank-two flavor symmetry and presented strong evidence that its infrared fixed point is the rank-one $\cN=3$ $\bZ_3$ S-fold SCFT. Our proposal is supported by matching the central charges and reproducing all known results for various limits of the superconformal index. Beyond these checks, our gauge-theory description yields, to our knowledge, the first gauge-theoretic expression for the full four-fugacity superconformal index of a genuine  and intrinsic $\cN=3$ SCFT, providing access to its refined protected information not captured by its known specializations.
Our construction also provides  an explicit realization of the   rank-one frozen   $\cN=2$ SCFT and its full index, which was inaccessible previously. 

This construction opens several avenues for exploring exotic strongly coupled theories through conventional gauge-theory techniques.

An immediate goal is to extend our construction to other $\cN=3$ SCFTs. A natural next target is the rank-one $\cN=3$ $\bZ_4$ S-fold SCFT with central charges $a=c=7/4$. Another important direction is to develop analogous constructions for non-Lagrangian $\cN=2$ S-fold SCFTs. Finding $\cN=1$ gauge theories that flow to these fixed points while retaining the relevant flavor symmetries would enable systematic calculations of their refined protected spectra. Such realizations could also clarify how the structures originally revealed by string-theoretic S-fold constructions emerge from four-dimensional gauge dynamics.

An explicit Lagrangian also offers a starting point for studying compactifications. In particular, it would be interesting to determine whether circle reduction leads to three-dimensional $\cN=2$ gauge theories flowing to $\cN=6$ SCFTs, potentially connecting four-dimensional supersymmetry enhancement to new three-dimensional gauge dualities.

More fundamentally, it remains to uncover the general principles governing supersymmetry enhancement along RG flows of $\cN=1$ gauge theories. In particular, our UV gauge theory has no origin from $\cN=2$, in contrast to \cite{Maruyoshi:2016tqk} . Understanding how additional supercharges emerge from strongly coupled dynamics could transform individual constructions into a systematic framework for discovering Lagrangian routes to non-Lagrangian fixed points.


\acknowledgements
We would like to thank  Gabi Zafrir  for discussions and comments on the draft.
This work was supported by the startup grant at SIMIS and the Shanghai Pujiang Program (No.~25PJA128).

 \bibliography{ref}

\begin{thebibliography}{25}%
\makeatletter
\providecommand \@ifxundefined [1]{%
 \@ifx{#1\undefined}
}%
\providecommand \@ifnum [1]{%
 \ifnum #1\expandafter \@firstoftwo
 \else \expandafter \@secondoftwo
 \fi
}%
\providecommand \@ifx [1]{%
 \ifx #1\expandafter \@firstoftwo
 \else \expandafter \@secondoftwo
 \fi
}%
\providecommand \natexlab [1]{#1}%
\providecommand \enquote  [1]{``#1''}%
\providecommand \bibnamefont  [1]{#1}%
\providecommand \bibfnamefont [1]{#1}%
\providecommand \citenamefont [1]{#1}%
\providecommand \href@noop [0]{\@secondoftwo}%
\providecommand \href [0]{\begingroup \@sanitize@url \@href}%
\providecommand \@href[1]{\@@startlink{#1}\@@href}%
\providecommand \@@href[1]{\endgroup#1\@@endlink}%
\providecommand \@sanitize@url [0]{\catcode `\\12\catcode `\$12\catcode
  `\&12\catcode `\#12\catcode `\^12\catcode `\_12\catcode `\%12\relax}%
\providecommand \@@startlink[1]{}%
\providecommand \@@endlink[0]{}%
\providecommand \url  [0]{\begingroup\@sanitize@url \@url }%
\providecommand \@url [1]{\endgroup\@href {#1}{\urlprefix }}%
\providecommand \urlprefix  [0]{URL }%
\providecommand \Eprint [0]{\href }%
\providecommand \doibase [0]{https://doi.org/}%
\providecommand \selectlanguage [0]{\@gobble}%
\providecommand \bibinfo  [0]{\@secondoftwo}%
\providecommand \bibfield  [0]{\@secondoftwo}%
\providecommand \translation [1]{[#1]}%
\providecommand \BibitemOpen [0]{}%
\providecommand \bibitemStop [0]{}%
\providecommand \bibitemNoStop [0]{.\EOS\space}%
\providecommand \EOS [0]{\spacefactor3000\relax}%
\providecommand \BibitemShut  [1]{\csname bibitem#1\endcsname}%
\let\auto@bib@innerbib\@empty
\bibitem [{\citenamefont {Aharony}\ and\ \citenamefont
  {Evtikhiev}(2016)}]{Aharony:2015oyb}%
  \BibitemOpen
  \bibfield  {author} {\bibinfo {author} {\bibfnamefont {O.}~\bibnamefont
  {Aharony}}\ and\ \bibinfo {author} {\bibfnamefont {M.}~\bibnamefont
  {Evtikhiev}},\ }\bibfield  {title} {\bibinfo {title} {{On four dimensional N
  = 3 superconformal theories}},\ }\href
  {https://doi.org/10.1007/JHEP04(2016)040} {\bibfield  {journal} {\bibinfo
  {journal} {JHEP}\ }\textbf {\bibinfo {volume} {04}},\ \bibinfo {pages}
  {040}},\ \Eprint {https://arxiv.org/abs/1512.03524} {arXiv:1512.03524
  [hep-th]} \BibitemShut {NoStop}%
\bibitem [{\citenamefont {Garc{\'\i}a-Etxebarria}\ and\ \citenamefont
  {Regalado}(2016)}]{Garcia-Etxebarria:2015wns}%
  \BibitemOpen
  \bibfield  {author} {\bibinfo {author} {\bibfnamefont {I.}~\bibnamefont
  {Garc{\'\i}a-Etxebarria}}\ and\ \bibinfo {author} {\bibfnamefont
  {D.}~\bibnamefont {Regalado}},\ }\bibfield  {title} {\bibinfo {title} {{$
  \mathcal{N}=3 $ four dimensional field theories}},\ }\href
  {https://doi.org/10.1007/JHEP03(2016)083} {\bibfield  {journal} {\bibinfo
  {journal} {JHEP}\ }\textbf {\bibinfo {volume} {03}},\ \bibinfo {pages}
  {083}},\ \Eprint {https://arxiv.org/abs/1512.06434} {arXiv:1512.06434
  [hep-th]} \BibitemShut {NoStop}%
\bibitem [{\citenamefont {Argyres}\ and\ \citenamefont
  {Douglas}(1995)}]{Argyres:1995jj}%
  \BibitemOpen
  \bibfield  {author} {\bibinfo {author} {\bibfnamefont {P.~C.}\ \bibnamefont
  {Argyres}}\ and\ \bibinfo {author} {\bibfnamefont {M.~R.}\ \bibnamefont
  {Douglas}},\ }\bibfield  {title} {\bibinfo {title} {{New phenomena in SU(3)
  supersymmetric gauge theory}},\ }\href
  {https://doi.org/10.1016/0550-3213(95)00281-V} {\bibfield  {journal}
  {\bibinfo  {journal} {Nucl. Phys. B}\ }\textbf {\bibinfo {volume} {448}},\
  \bibinfo {pages} {93} (\bibinfo {year} {1995})},\ \Eprint
  {https://arxiv.org/abs/hep-th/9505062} {arXiv:hep-th/9505062} \BibitemShut
  {NoStop}%
\bibitem [{\citenamefont {Gadde}\ \emph {et~al.}(2015)\citenamefont {Gadde},
  \citenamefont {Razamat},\ and\ \citenamefont {Willett}}]{Gadde:2015xta}%
  \BibitemOpen
  \bibfield  {author} {\bibinfo {author} {\bibfnamefont {A.}~\bibnamefont
  {Gadde}}, \bibinfo {author} {\bibfnamefont {S.~S.}\ \bibnamefont {Razamat}},\
  and\ \bibinfo {author} {\bibfnamefont {B.}~\bibnamefont {Willett}},\
  }\bibfield  {title} {\bibinfo {title} {{''Lagrangian'' for a Non-Lagrangian
  Field Theory with $\mathcal N=2$ Supersymmetry}},\ }\href
  {https://doi.org/10.1103/PhysRevLett.115.171604} {\bibfield  {journal}
  {\bibinfo  {journal} {Phys. Rev. Lett.}\ }\textbf {\bibinfo {volume} {115}},\
  \bibinfo {pages} {171604} (\bibinfo {year} {2015})},\ \Eprint
  {https://arxiv.org/abs/1505.05834} {arXiv:1505.05834 [hep-th]} \BibitemShut
  {NoStop}%
\bibitem [{\citenamefont {Maruyoshi}\ and\ \citenamefont
  {Song}(2017)}]{Maruyoshi:2016tqk}%
  \BibitemOpen
  \bibfield  {author} {\bibinfo {author} {\bibfnamefont {K.}~\bibnamefont
  {Maruyoshi}}\ and\ \bibinfo {author} {\bibfnamefont {J.}~\bibnamefont
  {Song}},\ }\bibfield  {title} {\bibinfo {title} {{Enhancement of
  Supersymmetry via Renormalization Group Flow and the Superconformal Index}},\
  }\href {https://doi.org/10.1103/PhysRevLett.118.151602} {\bibfield  {journal}
  {\bibinfo  {journal} {Phys. Rev. Lett.}\ }\textbf {\bibinfo {volume} {118}},\
  \bibinfo {pages} {151602} (\bibinfo {year} {2017})},\ \Eprint
  {https://arxiv.org/abs/1606.05632} {arXiv:1606.05632 [hep-th]} \BibitemShut
  {NoStop}%
\bibitem [{\citenamefont {Zafrir}(2021)}]{Zafrir:2020epd}%
  \BibitemOpen
  \bibfield  {author} {\bibinfo {author} {\bibfnamefont {G.}~\bibnamefont
  {Zafrir}},\ }\bibfield  {title} {\bibinfo {title} {{An $ \mathcal{N} $ = 1
  Lagrangian for an $ \mathcal{N} $ = 3 SCFT}},\ }\href
  {https://doi.org/10.1007/JHEP01(2021)062} {\bibfield  {journal} {\bibinfo
  {journal} {JHEP}\ }\textbf {\bibinfo {volume} {01}},\ \bibinfo {pages}
  {062}},\ \Eprint {https://arxiv.org/abs/2007.14955} {arXiv:2007.14955
  [hep-th]} \BibitemShut {NoStop}%
\bibitem [{Note1()}]{Note1}%
  \BibitemOpen
  \bibinfo {note} {\cite {Bourton:2018jwb} also computed the full index of an
  $\protect \mathcal {N}=3$ SCFT, which, however, arises from discrete gauging
  of an $\protect \mathcal {N}=4$ theory and is therefore not intrinsically
  $\protect \mathcal {N}=3$.}\BibitemShut {Stop}%
\bibitem [{\citenamefont {Argyres}\ \emph {et~al.}(2018)\citenamefont
  {Argyres}, \citenamefont {Lotito}, \citenamefont {L{\"u}},\ and\
  \citenamefont {Martone}}]{Argyres:2016xmc}%
  \BibitemOpen
  \bibfield  {author} {\bibinfo {author} {\bibfnamefont {P.}~\bibnamefont
  {Argyres}}, \bibinfo {author} {\bibfnamefont {M.}~\bibnamefont {Lotito}},
  \bibinfo {author} {\bibfnamefont {Y.}~\bibnamefont {L{\"u}}},\ and\ \bibinfo
  {author} {\bibfnamefont {M.}~\bibnamefont {Martone}},\ }\bibfield  {title}
  {\bibinfo {title} {{Geometric constraints on the space of $ \mathcal{N}$ = 2
  SCFTs. Part III: enhanced Coulomb branches and central charges}},\ }\href
  {https://doi.org/10.1007/JHEP02(2018)003} {\bibfield  {journal} {\bibinfo
  {journal} {JHEP}\ }\textbf {\bibinfo {volume} {02}},\ \bibinfo {pages}
  {003}},\ \Eprint {https://arxiv.org/abs/1609.04404} {arXiv:1609.04404
  [hep-th]} \BibitemShut {NoStop}%
\bibitem [{\citenamefont {Intriligator}\ and\ \citenamefont
  {Wecht}(2003)}]{Intriligator:2003jj}%
  \BibitemOpen
  \bibfield  {author} {\bibinfo {author} {\bibfnamefont {K.~A.}\ \bibnamefont
  {Intriligator}}\ and\ \bibinfo {author} {\bibfnamefont {B.}~\bibnamefont
  {Wecht}},\ }\bibfield  {title} {\bibinfo {title} {{The Exact superconformal R
  symmetry maximizes a}},\ }\href
  {https://doi.org/10.1016/S0550-3213(03)00459-0} {\bibfield  {journal}
  {\bibinfo  {journal} {Nucl. Phys. B}\ }\textbf {\bibinfo {volume} {667}},\
  \bibinfo {pages} {183} (\bibinfo {year} {2003})},\ \Eprint
  {https://arxiv.org/abs/hep-th/0304128} {arXiv:hep-th/0304128} \BibitemShut
  {NoStop}%
\bibitem [{\citenamefont {Gadde}(2022)}]{Gadde:2020yah}%
  \BibitemOpen
  \bibfield  {author} {\bibinfo {author} {\bibfnamefont {A.}~\bibnamefont
  {Gadde}},\ }\bibfield  {title} {\bibinfo {title} {{Lectures on the
  Superconformal Index}},\ }\href {https://doi.org/10.1088/1751-8121/ac42ac}
  {\bibfield  {journal} {\bibinfo  {journal} {J. Phys. A}\ }\textbf {\bibinfo
  {volume} {55}},\ \bibinfo {pages} {063001} (\bibinfo {year} {2022})},\
  \Eprint {https://arxiv.org/abs/2006.13630} {arXiv:2006.13630 [hep-th]}
  \BibitemShut {NoStop}%
\bibitem [{Note2()}]{Note2}%
  \BibitemOpen
  \bibinfo {note} {Because of the subtlety associated with $R(Y)=0$, the
  contour integral should be evaluated in the chamber $|s|=|\protect \sqrt {\xi
  /\sigma ^3}|<1$, with the limit $s\to 1$ taken afterward if needed. For
  convenience, one may rescale the fugacities as $\xi =\tau ^{1/4}\protect \hat
  \xi $ and $\sigma =\tau ^{-1/4}\protect \hat \sigma $. We then use $\tau
  ,\rho ,\protect \hat \xi ,\protect \hat \sigma $ as independent variables and
  expand the integrand in $\tau $. The integral can be evaluated order by order
  by taking residues at the origins of the gauge-fugacity planes, and the final
  result can be reexpressed in terms of $\tau ,\rho ,\sigma ,\xi
  $.}\BibitemShut {Stop}%
\bibitem [{Note3()}]{Note3}%
  \BibitemOpen
  \bibinfo {note} {We have evaluated the full index up to the order $\tau
  ^{60}$ and we show the result up to order $\tau ^{20}$ in
  appendix.}\BibitemShut {Stop}%
\bibitem [{\citenamefont {Evtikhiev}(2018)}]{Evtikhiev:2017heo}%
  \BibitemOpen
  \bibfield  {author} {\bibinfo {author} {\bibfnamefont {M.}~\bibnamefont
  {Evtikhiev}},\ }\bibfield  {title} {\bibinfo {title} {{Studying
  superconformal symmetry enhancement through indices}},\ }\href
  {https://doi.org/10.1007/JHEP04(2018)120} {\bibfield  {journal} {\bibinfo
  {journal} {JHEP}\ }\textbf {\bibinfo {volume} {04}},\ \bibinfo {pages}
  {120}},\ \Eprint {https://arxiv.org/abs/1708.08307} {arXiv:1708.08307
  [hep-th]} \BibitemShut {NoStop}%
\bibitem [{Note4()}]{Note4}%
  \BibitemOpen
  \bibinfo {note} {The general statement in \cite {Evtikhiev:2017heo} is that:
  for an interacting theory whose reduced index is $ (1-p)(1-q)(\protect
  \mathcal I-1)$, a positive integer coefficient \(k\) of \(\tau ^7\chi _2(\rho
  )\) implies enhancement of supersymmetry to at least \(\protect \mathcal
  N=k+1\). Here we use the flavor resolved version of this statement. Also it
  is easy to argue that our IR theory can not have $\protect \mathcal {N}=4$
  supersymmetry.}\BibitemShut {Stop}%
\bibitem [{\citenamefont {Cordova}\ \emph {et~al.}(2019)\citenamefont
  {Cordova}, \citenamefont {Dumitrescu},\ and\ \citenamefont
  {Intriligator}}]{Cordova:2016emh}%
  \BibitemOpen
  \bibfield  {author} {\bibinfo {author} {\bibfnamefont {C.}~\bibnamefont
  {Cordova}}, \bibinfo {author} {\bibfnamefont {T.~T.}\ \bibnamefont
  {Dumitrescu}},\ and\ \bibinfo {author} {\bibfnamefont {K.}~\bibnamefont
  {Intriligator}},\ }\bibfield  {title} {\bibinfo {title} {{Multiplets of
  Superconformal Symmetry in Diverse Dimensions}},\ }\href
  {https://doi.org/10.1007/JHEP03(2019)163} {\bibfield  {journal} {\bibinfo
  {journal} {JHEP}\ }\textbf {\bibinfo {volume} {03}},\ \bibinfo {pages}
  {163}},\ \Eprint {https://arxiv.org/abs/1612.00809} {arXiv:1612.00809
  [hep-th]} \BibitemShut {NoStop}%
\bibitem [{\citenamefont {Beem}\ \emph {et~al.}(2015)\citenamefont {Beem},
  \citenamefont {Lemos}, \citenamefont {Liendo}, \citenamefont {Peelaers},
  \citenamefont {Rastelli},\ and\ \citenamefont {van Rees}}]{Beem:2013sza}%
  \BibitemOpen
  \bibfield  {author} {\bibinfo {author} {\bibfnamefont {C.}~\bibnamefont
  {Beem}}, \bibinfo {author} {\bibfnamefont {M.}~\bibnamefont {Lemos}},
  \bibinfo {author} {\bibfnamefont {P.}~\bibnamefont {Liendo}}, \bibinfo
  {author} {\bibfnamefont {W.}~\bibnamefont {Peelaers}}, \bibinfo {author}
  {\bibfnamefont {L.}~\bibnamefont {Rastelli}},\ and\ \bibinfo {author}
  {\bibfnamefont {B.~C.}\ \bibnamefont {van Rees}},\ }\bibfield  {title}
  {\bibinfo {title} {{Infinite Chiral Symmetry in Four Dimensions}},\ }\href
  {https://doi.org/10.1007/s00220-014-2272-x} {\bibfield  {journal} {\bibinfo
  {journal} {Commun. Math. Phys.}\ }\textbf {\bibinfo {volume} {336}},\
  \bibinfo {pages} {1359} (\bibinfo {year} {2015})},\ \Eprint
  {https://arxiv.org/abs/1312.5344} {arXiv:1312.5344 [hep-th]} \BibitemShut
  {NoStop}%
\bibitem [{\citenamefont {Nishinaka}\ and\ \citenamefont
  {Tachikawa}(2016)}]{Nishinaka:2016hbw}%
  \BibitemOpen
  \bibfield  {author} {\bibinfo {author} {\bibfnamefont {T.}~\bibnamefont
  {Nishinaka}}\ and\ \bibinfo {author} {\bibfnamefont {Y.}~\bibnamefont
  {Tachikawa}},\ }\bibfield  {title} {\bibinfo {title} {{On 4d rank-one $
  \mathcal{N}=3 $ superconformal field theories}},\ }\href
  {https://doi.org/10.1007/JHEP09(2016)116} {\bibfield  {journal} {\bibinfo
  {journal} {JHEP}\ }\textbf {\bibinfo {volume} {09}},\ \bibinfo {pages}
  {116}},\ \Eprint {https://arxiv.org/abs/1602.01503} {arXiv:1602.01503
  [hep-th]} \BibitemShut {NoStop}%
\bibitem [{\citenamefont {Bonetti}\ \emph {et~al.}(2019)\citenamefont
  {Bonetti}, \citenamefont {Meneghelli},\ and\ \citenamefont
  {Rastelli}}]{Bonetti:2018fqz}%
  \BibitemOpen
  \bibfield  {author} {\bibinfo {author} {\bibfnamefont {F.}~\bibnamefont
  {Bonetti}}, \bibinfo {author} {\bibfnamefont {C.}~\bibnamefont
  {Meneghelli}},\ and\ \bibinfo {author} {\bibfnamefont {L.}~\bibnamefont
  {Rastelli}},\ }\bibfield  {title} {\bibinfo {title} {{VOAs labelled by
  complex reflection groups and 4d SCFTs}},\ }\href
  {https://doi.org/10.1007/JHEP05(2019)155} {\bibfield  {journal} {\bibinfo
  {journal} {JHEP}\ }\textbf {\bibinfo {volume} {05}},\ \bibinfo {pages}
  {155}},\ \Eprint {https://arxiv.org/abs/1810.03612} {arXiv:1810.03612
  [hep-th]} \BibitemShut {NoStop}%
\bibitem [{\citenamefont {Jiang}()}]{Jiang2026}%
  \BibitemOpen
  \bibfield  {author} {\bibinfo {author} {\bibfnamefont {H.}~\bibnamefont
  {Jiang}},\ }\bibfield  {title} {\bibinfo {title} {in preparation},\
  }\href@noop {} {\ }\BibitemShut {NoStop}%
\bibitem [{\citenamefont {Guo}\ \emph {et~al.}(2026)\citenamefont {Guo},
  \citenamefont {Nawata}, \citenamefont {Pan},\ and\ \citenamefont
  {Zhang}}]{Guo:2026lsl}%
  \BibitemOpen
  \bibfield  {author} {\bibinfo {author} {\bibfnamefont {Z.}~\bibnamefont
  {Guo}}, \bibinfo {author} {\bibfnamefont {S.}~\bibnamefont {Nawata}},
  \bibinfo {author} {\bibfnamefont {Y.}~\bibnamefont {Pan}},\ and\ \bibinfo
  {author} {\bibfnamefont {Q.}~\bibnamefont {Zhang}},\ }\bibfield  {title}
  {\bibinfo {title} {{Rank-one 4d $\mathcal N=3$ SCFTs: Schur index, VOA
  modules, and modularity}},\ }\href@noop {} {\  (\bibinfo {year} {2026})},\
  \Eprint {https://arxiv.org/abs/2609.10685} {arXiv:2609.10685 [hep-th]}
  \BibitemShut {NoStop}%
\bibitem [{\citenamefont {Agarwal}\ \emph {et~al.}(2021)\citenamefont
  {Agarwal}, \citenamefont {Andriolo}, \citenamefont {K{\'a}ntor},\ and\
  \citenamefont {Papageorgakis}}]{Agarwal:2021oyl}%
  \BibitemOpen
  \bibfield  {author} {\bibinfo {author} {\bibfnamefont {P.}~\bibnamefont
  {Agarwal}}, \bibinfo {author} {\bibfnamefont {E.}~\bibnamefont {Andriolo}},
  \bibinfo {author} {\bibfnamefont {G.}~\bibnamefont {K{\'a}ntor}},\ and\
  \bibinfo {author} {\bibfnamefont {C.}~\bibnamefont {Papageorgakis}},\
  }\bibfield  {title} {\bibinfo {title} {{Macdonald indices for
  four-dimensional $\mathcal{N}$=3 theories}},\ }\href
  {https://doi.org/10.1103/PhysRevD.103.L121701} {\bibfield  {journal}
  {\bibinfo  {journal} {Phys. Rev. D}\ }\textbf {\bibinfo {volume} {103}},\
  \bibinfo {pages} {L121701} (\bibinfo {year} {2021})},\ \Eprint
  {https://arxiv.org/abs/2103.00985} {arXiv:2103.00985 [hep-th]} \BibitemShut
  {NoStop}%
\bibitem [{\citenamefont {Feigin}\ \emph {et~al.}(2007)\citenamefont {Feigin},
  \citenamefont {Feigin},\ and\ \citenamefont {Tipunin}}]{Feigin:2007sp}%
  \BibitemOpen
  \bibfield  {author} {\bibinfo {author} {\bibfnamefont {B.}~\bibnamefont
  {Feigin}}, \bibinfo {author} {\bibfnamefont {E.}~\bibnamefont {Feigin}},\
  and\ \bibinfo {author} {\bibfnamefont {I.}~\bibnamefont {Tipunin}},\
  }\bibfield  {title} {\bibinfo {title} {{Fermionic formulas for (1,p)
  logarithmic model characters in Phi{2,1} quasiparticle realisation}},\
  }\href@noop {} {\  (\bibinfo {year} {2007})},\ \Eprint
  {https://arxiv.org/abs/0704.2464} {arXiv:0704.2464 [hep-th]} \BibitemShut
  {NoStop}%
\bibitem [{\citenamefont {Feigin}\ and\ \citenamefont
  {Tipunin}(2008)}]{feigin2008characters}%
  \BibitemOpen
  \bibfield  {author} {\bibinfo {author} {\bibfnamefont {B.~L.}\ \bibnamefont
  {Feigin}}\ and\ \bibinfo {author} {\bibfnamefont {I.~Y.}\ \bibnamefont
  {Tipunin}},\ }\href@noop {} {\bibinfo {title} {Characters of coinvariants in
  (1,p) logarithmic models}} (\bibinfo {year} {2008}),\ \Eprint
  {https://arxiv.org/abs/0805.4096} {arXiv:0805.4096 [math.QA]} \BibitemShut
  {NoStop}%
\bibitem [{\citenamefont {Bourton}\ \emph {et~al.}(2018)\citenamefont
  {Bourton}, \citenamefont {Pini},\ and\ \citenamefont
  {Pomoni}}]{Bourton:2018jwb}%
  \BibitemOpen
  \bibfield  {author} {\bibinfo {author} {\bibfnamefont {T.}~\bibnamefont
  {Bourton}}, \bibinfo {author} {\bibfnamefont {A.}~\bibnamefont {Pini}},\ and\
  \bibinfo {author} {\bibfnamefont {E.}~\bibnamefont {Pomoni}},\ }\bibfield
  {title} {\bibinfo {title} {{4d $\mathcal{N}=3$ indices via discrete
  gauging}},\ }\href {https://doi.org/10.1007/JHEP10(2018)131} {\bibfield
  {journal} {\bibinfo  {journal} {JHEP}\ }\textbf {\bibinfo {volume} {10}},\
  \bibinfo {pages} {131}},\ \Eprint {https://arxiv.org/abs/1804.05396}
  {arXiv:1804.05396 [hep-th]} \BibitemShut {NoStop}%
\bibitem [{\citenamefont
  {Rains}(2005)}]{rains2005transformationselliptichypergometricintegrals}%
  \BibitemOpen
  \bibfield  {author} {\bibinfo {author} {\bibfnamefont {E.~M.}\ \bibnamefont
  {Rains}},\ }\href {https://arxiv.org/abs/math/0309252} {\bibinfo {title}
  {Transformations of elliptic hypergometric integrals}} (\bibinfo {year}
  {2005}),\ \Eprint {https://arxiv.org/abs/math/0309252} {arXiv:math/0309252
  [math.QA]} \BibitemShut {NoStop}%
\end{thebibliography}%
 
\onecolumngrid
 
 
 \appendix
 \section{Perturbative Expansion of Full Index}\label{appPerIndx}

We can use $\tau,\rho,\sigma,\xi$ as fugacities of the index and expand in $\tau$
\be
\cI(\tau,\rho,\sigma,\xi)=\sum_{n=0}^{\infty}C_n(\rho,\sigma,\xi)  \tau^n 
\ee
where the coefficients $C_n $ are functions of the remaining fugacities. We have computed $C_n$ for all $n\le 60$.  Below we present the results for $n\le 20$. 
For   convenience, we use the following notation 
\be
X_d=\chi_d(\xi),\qquad R_d=\chi_d(\rho),\qquad
\chi_d(w)=\sum_{k=0}^{d-1}w^{d-1-2k},\qquad X_1=R_1=1.
\ee
Then the coefficients through order $\tau^{20}$  are  $   C_0=1,  C_1=C_2=C_3=0$ and 
\allowdisplaybreaks
 \begin{flalign*}
C_{4}&={}X_{2}\sigma^{-1}, && \\[2pt]
C_{5}&={}-R_{2}\sigma^{-2}, && \\[2pt]
C_{6}&={}\sigma^{-6} {}-1 {}-X_{3} {}+X_{4}\sigma^{3}, && \\[2pt]
C_{7}&={}2 X_{2}\,R_{2}\sigma^{-1} {}-X_{3}\,R_{2}\sigma^{2}, && \\[2pt]
C_{8}&={}-X_{2}\sigma^{-5} {}-\bigl(1 {}+R_{3} {}-X_{3}\bigr)\sigma^{-2} {}+2 X_{2}\sigma {}-X_{3}\sigma^{4}, && \\[2pt]
C_{9}&={}R_{2}\sigma^{-6} {}-X_{2}\,R_{2}\sigma^{-3} {}-2 R_{2} {}-X_{3}\,R_{2} {}+\bigl(X_{2}\,R_{2} {}+X_{4}\,R_{2}\bigr)\sigma^{3}, && \\[2pt]
C_{10}&={}X_{2}\sigma^{-7} {}+2\sigma^{-4} {}-\bigl(X_{2} {}-2 X_{2}\,R_{3} {}+X_{4}\bigr)\sigma^{-1} {}-\bigl(1 {}+X_{3} {}+X_{3}\,R_{3} {}-X_{5}\bigr)\sigma^{2}, && \\[2pt]
C_{11}&={}-R_{2}\sigma^{-8} {}-X_{2}\,R_{2}\sigma^{-5} {}+\bigl(2 R_{2} {}-R_{4} {}+3 X_{3}\,R_{2}\bigr)\sigma^{-2} {}+\bigl(2 X_{2}\,R_{2} {}-X_{4}\,R_{2}\bigr)\sigma {}-X_{3}\,R_{2}\sigma^{4}, && \\[2pt]
C_{12}&={}\sigma^{-12} {}-\bigl(2 {}-R_{3} {}+X_{3}\bigr)\sigma^{-6} {}-\bigl(3 X_{2} {}+3 X_{2}\,R_{3} {}-X_{4}\bigr)\sigma^{-3} {}-2 R_{3} {}+4 X_{3} {}-X_{3}\,R_{3} {}+\bigl(X_{2}\,R_{3} {}-2 X_{4} {}+X_{4}\,R_{3}  && \\*
&\quad {}-X_{6}\bigr)\sigma^{3} {}+X_{7}\sigma^{6}, && \\[2pt]
C_{13}&={}3 X_{2}\,R_{2}\sigma^{-7} {}+\bigl(3 R_{2} {}+R_{4} {}-X_{3}\,R_{2}\bigr)\sigma^{-4} {}-\bigl(7 X_{2}\,R_{2} {}-2 X_{2}\,R_{4} {}+3 X_{4}\,R_{2}\bigr)\sigma^{-1} {}+\bigl(3 X_{3}\,R_{2} {}-X_{3}\,R_{4} && \\*
&\quad {}+3 X_{5}\,R_{2}\bigr)\sigma^{2} {}-X_{6}\,R_{2}\sigma^{5}, && \\[2pt]
C_{14}&={}-X_{2}\sigma^{-11} {}-\bigl(1 {}+2 R_{3} {}-X_{3}\bigr)\sigma^{-8} {}+\bigl(4 X_{2} {}-X_{2}\,R_{3}\bigr)\sigma^{-5} {}+\bigl(7 {}+5 R_{3} {}-R_{5} {}+2 X_{3} {}+7 X_{3}\,R_{3} {}-X_{5}\bigr)\sigma^{-2} && \\*
&\quad {}-\bigl(6 X_{2} {}+5 X_{4} {}+3 X_{4}\,R_{3} {}-X_{6}\bigr)\sigma {}+\bigl(1 {}+2 X_{3} {}-X_{3}\,R_{3} {}+3 X_{5}\bigr)\sigma^{4} {}-X_{6}\sigma^{7}, && \\[2pt]
C_{15}&={}R_{2}\sigma^{-12} {}-X_{2}\,R_{2}\sigma^{-9} {}-\bigl(6 R_{2} {}-R_{4} {}+3 X_{3}\,R_{2}\bigr)\sigma^{-6} {}-\bigl(5 X_{2}\,R_{2} {}+5 X_{2}\,R_{4} {}-3 X_{4}\,R_{2}\bigr)\sigma^{-3} {}+7 R_{2} {}-2 R_{4} && \\*
&\quad {}+11 X_{3}\,R_{2} {}-\bigl(5 X_{2}\,R_{2} {}-X_{2}\,R_{4} {}+8 X_{4}\,R_{2} {}-X_{4}\,R_{4} {}+2 X_{6}\,R_{2}\bigr)\sigma^{3} {}+\bigl(2 X_{5}\,R_{2} {}+X_{7}\,R_{2}\bigr)\sigma^{6}, && \\[2pt]
C_{16}&={}X_{2}\sigma^{-13} {}+3\sigma^{-10} {}+\bigl(X_{2} {}+6 X_{2}\,R_{3} {}-X_{4}\bigr)\sigma^{-7} {}-\bigl(4 {}-4 R_{3} {}-R_{5} {}+4 X_{3} {}+3 X_{3}\,R_{3} {}-X_{5}\bigr)\sigma^{-4} {}-\bigl(10 X_{2} && \\*
&\quad {}+15 X_{2}\,R_{3} {}-2 X_{2}\,R_{5} {}-X_{4} {}+5 X_{4}\,R_{3}\bigr)\sigma^{-1} {}+\bigl(5 {}+4 R_{3} {}+13 X_{3} {}+10 X_{3}\,R_{3} {}-X_{3}\,R_{5} {}+2 X_{5} {}+6 X_{5}\,R_{3} && \\*
&\quad {}-X_{7}\bigr)\sigma^{2} {}-\bigl(X_{2} {}+3 X_{4} {}+X_{4}\,R_{3} {}+X_{6} {}+2 X_{6}\,R_{3} {}-X_{8}\bigr)\sigma^{5}, && \\[2pt]
C_{17}&={}-R_{2}\sigma^{-14} {}-2 X_{2}\,R_{2}\sigma^{-11} {}-\bigl(3 R_{4} {}-3 X_{3}\,R_{2}\bigr)\sigma^{-8} {}+11 X_{2}\,R_{2}\sigma^{-5} {}+\bigl(11 R_{2} {}+10 R_{4} {}-R_{6} {}+9 X_{3}\,R_{4} && \\*
&\quad {}-3 X_{5}\,R_{2}\bigr)\sigma^{-2} {}-\bigl(21 X_{2}\,R_{2} {}+4 X_{2}\,R_{4} {}+10 X_{4}\,R_{2} {}+6 X_{4}\,R_{4} {}-3 X_{6}\,R_{2}\bigr)\sigma {}+\bigl(3 R_{2} {}+8 X_{3}\,R_{2} {}-X_{3}\,R_{4} && \\*
&\quad {}+6 X_{5}\,R_{2} {}+X_{5}\,R_{4} {}-X_{7}\,R_{2}\bigr)\sigma^{4} {}-\bigl(X_{4}\,R_{2} {}+2 X_{6}\,R_{2}\bigr)\sigma^{7}, && \\[2pt]
C_{18}&={}\sigma^{-18} {}-\bigl(2 {}-2 R_{3} {}+X_{3}\bigr)\sigma^{-12} {}-\bigl(4 X_{2} {}+3 X_{2}\,R_{3} {}-X_{4}\bigr)\sigma^{-9} {}-\bigl(5 {}+12 R_{3} {}-R_{5} {}-X_{3} {}+6 X_{3}\,R_{3}\bigr)\sigma^{-6} {}+\bigl(8 X_{2} && \\*
&\quad {}-7 X_{2}\,R_{5} {}+3 X_{4} {}+9 X_{4}\,R_{3} {}-X_{6}\bigr)\sigma^{-3} {}+11 {}+14 R_{3} {}-R_{5} {}+4 X_{3} {}+21 X_{3}\,R_{3} {}+X_{3}\,R_{5} {}-3 X_{5} {}-2 X_{5}\,R_{3} && \\*
&\quad {}+X_{7} {}-\bigl(10 X_{2} {}+13 X_{2}\,R_{3} {}-X_{2}\,R_{5} {}+8 X_{4} {}+16 X_{4}\,R_{3} {}-X_{4}\,R_{5} {}-3 X_{6} {}+3 X_{6}\,R_{3}\bigr)\sigma^{3} {}+\bigl(X_{3} {}+2 X_{3}\,R_{3} && \\*
&\quad {}+3 X_{5} {}+4 X_{5}\,R_{3} {}-2 X_{7} {}+2 X_{7}\,R_{3} {}-X_{9}\bigr)\sigma^{6} {}+X_{10}\sigma^{9}, && \\[2pt]
C_{19}&={}3 X_{2}\,R_{2}\sigma^{-13} {}+\bigl(5 R_{2} {}+R_{4}\bigr)\sigma^{-10} {}-\bigl(2 X_{2}\,R_{2} {}-9 X_{2}\,R_{4} {}+3 X_{4}\,R_{2}\bigr)\sigma^{-7} {}-\bigl(14 R_{2} {}-R_{4} {}-2 R_{6} {}+13 X_{3}\,R_{2} && \\*
&\quad {}+9 X_{3}\,R_{4} {}-3 X_{5}\,R_{2}\bigr)\sigma^{-4} {}-\bigl(11 X_{2}\,R_{2} {}+24 X_{2}\,R_{4} {}-2 X_{2}\,R_{6} {}-10 X_{4}\,R_{2} {}+6 X_{4}\,R_{4}\bigr)\sigma^{-1} {}+\bigl(17 R_{2} {}+8 R_{4} && \\*
&\quad {}+26 X_{3}\,R_{2} {}+19 X_{3}\,R_{4} {}-X_{3}\,R_{6} {}-2 X_{5}\,R_{2} {}+9 X_{5}\,R_{4} {}-3 X_{7}\,R_{2}\bigr)\sigma^{2} {}-\bigl(4 X_{2}\,R_{2} {}+X_{2}\,R_{4} {}+8 X_{4}\,R_{2} {}+3 X_{4}\,R_{4} && \\*
&\quad {}+3 X_{6}\,R_{4} {}-3 X_{8}\,R_{2}\bigr)\sigma^{5} {}+\bigl(X_{5}\,R_{2} {}-X_{9}\,R_{2}\bigr)\sigma^{8}, && \\[2pt]
C_{20}&={}-X_{2}\sigma^{-17} {}-\bigl(1 {}+2 R_{3} {}-X_{3}\bigr)\sigma^{-14} {}+\bigl(3 X_{2} {}-3 X_{2}\,R_{3}\bigr)\sigma^{-11} {}+\bigl(8 {}+4 R_{3} {}-4 R_{5} {}+2 X_{3} {}+9 X_{3}\,R_{3} {}-X_{5}\bigr)\sigma^{-8} && \\*
&\quad {}+\bigl(5 X_{2} {}+24 X_{2}\,R_{3} {}+2 X_{2}\,R_{5} {}-3 X_{4} {}-X_{4}\,R_{3} {}+X_{6}\bigr)\sigma^{-5} {}-\bigl(9 {}-7 R_{3} {}-13 R_{5} {}+R_{7} {}+16 X_{3} {}+19 X_{3}\,R_{3} && \\*
&\quad {}-13 X_{3}\,R_{5} {}+X_{5} {}+9 X_{5}\,R_{3}\bigr)\sigma^{-2} {}-\bigl(12 X_{2} {}+35 X_{2}\,R_{3} {}+8 X_{2}\,R_{5} {}-9 X_{4} {}+9 X_{4}\,R_{3} {}+9 X_{4}\,R_{5} {}-3 X_{6} && \\*
&\quad {}-9 X_{6}\,R_{3} {}+X_{8}\bigr)\sigma {}+\bigl(1 {}+7 R_{3} {}+6 X_{3} {}+16 X_{3}\,R_{3} {}-4 X_{5} {}+10 X_{5}\,R_{3} {}+X_{5}\,R_{5} {}-4 X_{7} {}-3 X_{7}\,R_{3} {}+X_{9}\bigr)\sigma^{4} && \\*
&\quad {}-\bigl(X_{2}\,R_{3} {}+2 X_{4} {}+3 X_{4}\,R_{3} {}+3 X_{6}\,R_{3} {}-3 X_{8}\bigr)\sigma^{7} {}-X_{9}\sigma^{10}. && \\[2pt]
\end{flalign*}

  \section{Coulomb index}\label{appCBIndx}

We now evaluate   the Coulomb index analytically, which corresponds to the limit $p,q\to 0$ with $u\equiv pq/t$ fixed. To realize this limit, 
 we introduce $\tau^6=pq, t=\tau^6/u$ in the limit $\tau\to 0$ with fixed $u,s$.  
   
 Then the gauge  integral for the index can be written as
   \begin{align}
\mathcal{I} 
&=  
\oint \frac{dx}{2\pi i x}\frac{dy}{2\pi i y}\frac{dz}{2\pi i z} \frac{\Gamma_e\!\left( Lz^{\pm 1}\right)}{\Gamma_e(z^{\pm 2})}F(z;x,y)
   \end{align}
   with
   \begin{align}\label{Findts}
F(z;x,y)&=
\left(\frac{\kappa}{2}\right)^3
\frac{1}{\Gamma_e(x^{\pm 2})\Gamma_e(y^{\pm 2}) }
\Gamma_e\!\left(s\,\tau^3 u^{-1/2}\right)
\Gamma_e\!\left(s\,\tau^3 u^{-1/2} z^{\pm 2}\right)  \nonumber \\[4pt]
&\quad\times
\Gamma_e\!\left(\tau^{3/2}s^{-1/2}u^{1/4}z^{\pm 1}y^{\pm 1}x^{\pm 1}\right)
\Gamma_e\!\left(\tau^{3/2}s^{-1/2}u^{-1/4}z^{\pm 1}y^{\pm 1}\right) \Gamma_e\!\left(s\,y^{\pm 1}x^{\pm 1}\right)
\nonumber \\[4pt]&\quad\times
\Gamma_e\!\left(\tau^{9/2}s^{-1/2}u^{-1/4}z^{\pm 1}\right)
\Gamma_e\!\left(\tau^6 s\,u^{-3/2}y^{\pm 1}\right)
\Gamma_e\!\left(\tau^3 u\,y^{\pm 1}\right)
\nonumber\\[4pt]&\quad\times
\Gamma_e\!\left(\tau^3 s\,x^{\pm 1}\right)   
\Gamma_e\!\left(\tau^3 s^{-1}u^{-1}x^{\pm 1}\right)
\Gamma_e\!\left(\tau^3 s\,u^{1/2}\right)
\Gamma_e\!\left(\tau^6 s^{-2}\right),
\end{align}
where    $L=\tau^{-3/2}s^{-1/2}u^{7/4}$. Note that all the elliptic gamma factors  have non-negative powers of $\tau$, except for $\Gamma_e\!\left( L z^{\pm 1}\right)=\Gamma_e(\tau^{-3/2}s^{-1/2}u^{7/4}z^{\pm 1} )$.

In the limit $\tau \to 0$, we have the following simplification 
\begin{align}\label{simptau}
\kappa \to 1, \qquad
\Gamma_e(w)  \to \frac{1}{1-w}, \qquad
\Gamma_e(\tau^6 w)  \to 1 - w^{-1},  \qquad
\Gamma_e(\tau^{ \alpha} w)  \to 1,   \qquad 0 < \alpha < 6.
\end{align}
and 
\[
\lim_{w\to 1}(1-w)\Gamma_e(w)=\kappa^{-1},
\qquad\qquad
\Gamma_e(w)\Gamma_e(\tau^6/w) = 1.
\]

To perform the $z$ integral, we can consider the following more general type of integral
\be
J_\cC[F] =
 \oint_\cC\frac{dz}{2\pi iz} \frac{\Gamma_e (Lz^{\pm 1})}{\Gamma_e ( z^{\pm 2})}F(z;x,y)
\ee
where $F(z;x,y)=F(1/z;x,y)$ is regular at $L^{\pm 1}$ and the contour integral along $\cC$ is chosen such that
\be
J_\cC[F]=J_\bT[F]+\text{Res}_{z=L} \cF-\text{Res}_{z=1/L} \cF,\qquad
 \cF(z;x,y)= \frac{1}{z}\frac{\Gamma_e (Lz^{\pm 1})}{\Gamma_e ( z^{\pm 2})}F(z;x,y)
\ee
where $\bT$ is the unit circle.  The contour difference adds the first residue and subtracts the second. This contour prescription implements the meromorphic continuation in \(L\), away from contour-pinching singularities   \cite{rains2005transformationselliptichypergometricintegrals}. 

The residue can be computed
\begin{equation*}
\operatorname*{Res}_{z=L} \cF = \frac{F(L; x, y)}{\kappa \Gamma_e(L^{-2})}, \qquad \operatorname*{Res}_{z=L^{-1}} \cF= -\frac{F(L^{-1}; x, y)}{\kappa \Gamma_e(L^{-2})}.
\end{equation*}
 Consequently,
\begin{equation} 
J_{\mathcal{C}}[F] = J_{\mathbb{T}}[F] + \frac{F(L; x, y) + F(L^{-1}; x, y)}{\kappa\Gamma_e(L^{-2})} = J_{\mathbb{T}}[F] + 
 \frac{2F(L; x, y)}{\kappa\Gamma_e(  L^{-2}) } .
\end{equation}

One can show that the    unit-circle contribution $J_{\mathbb{T}}[F]$  vanishes  in the limit $\tau\to 0$ after performing the remaining integrations, essentially due to the fact that $\Gamma_e (Lz^{\pm 1})\sim L^{-2}\propto  \tau^3$ for $|z|=1$. Therefore   we will only focus on the second term above. In the limit $\tau\to 0$, we can simplify   \eqref{Findts} using  \eqref{simptau} and obtain 

\beqn
\lim_{\tau\to 0} \cI&=&\lim_{\tau\to 0} \oint \frac{dx}{2\pi i x}\frac{dy}{2\pi i y} \frac{2F(L; x, y)}{\kappa\Gamma_e(  L^{-2}) } 
\\&=& 
 \frac{(1-s^2)(1-u^2)(1-u^4/s^2)}{1-u^3} \oint d\mu(x) d\mu(y) K_s(x,y) K_{u^2/s}(x,y)
 \label{ImKK}
\eeqn
where
\be
K_v(x,y) = \prod_{\epsilon,\eta=\pm1} \frac{1}{  1 - v x^{\epsilon}y^{\eta}  },
\qquad
\mathrm{d}\mu(x) = \frac{1}{2}\frac{\mathrm{d}x}{2\pi i x}
(1-x^2)(1-x^{-2}),
\qquad
\mathrm{d}\mu(y) = \frac{1}{2}\frac{\mathrm{d}y}{2\pi i y}
(1-y^2)(1-y^{-2}).\qquad
\ee

For \(|v|<1\), the \(SU(2)\) Cauchy identity reads
\[
K_v(x,y)=\frac{1}{1-v^2}\sum_{n=0}^{\infty}v^n\chi_{n+1}(x)\chi_{n+1}(y),
\qquad
\chi_{n+1}(x)=x^n+x^{n-2}+\cdots+x^{-n}.
\]
Using the orthonormal property of the character \(\int \mathrm{d}\mu(x)\chi_{n+1}(x)\chi_{m+1}(x)=\delta_{nm}\), the absolutely convergent series gives
\[
\begin{aligned}
\int \mathrm{d}\mu(x)\,\mathrm{d}\mu(y)\,K_s(x,y)K_{u^2/s}(x,y)
&=
\frac{\sum_{n= 0}^\infty u^{2n}}{(1-s^2)(1-u^4/s^2)}
 =
\frac{1}{(1-s^2)(1-u^4/s^2)(1-u^2)}.
\end{aligned}
\]
As a result 
\be
 \cI_{C}
=
 \frac{(1-s^2)(1-u^2)(1-u^4/s^2)}{1-u^3}  \frac{1}{(1-s^2)(1-u^4/s^2)(1-u^2)}
 =
 \frac{1}{1-u^3}.
\ee
The same result can also be obtained by applying the residue theorem directly to the integral in Eq. \eqref{ImKK} by accounting for the  contributions from  the relevant poles noting that $|s|,|u^2/s|<1$. The above result  can also be verified through the Coulomb limit of our perturbative  series expansion of the full index.

  \end{document}